\documentclass[10pt]{iopart}
\usepackage{graphicx}
\usepackage{lineno}
\usepackage{stfloats}
\usepackage{multicol}
\usepackage{array}
\usepackage{verbatim}

\begin{document}

\begin{flushleft}
\review[]{Recent advances in InGaAs/InP single-photon detectors}

\author{Chao Yu$^{1,2}$, Qi Xu$^{1,2}$, and Jun Zhang$^{1,2,3}$}

\address{$^1$ Hefei National Research Center for Physical Sciences at the Microscale and School of Physical Sciences, University of Science and Technology of China, Hefei 230026, China}
\address{$^2$ CAS Center for Excellence in Quantum Information and Quantum Physics, University of Science and Technology of China, Hefei 230026, China}
\address{$^3$ Hefei National Laboratory, University of Science and Technology of China, Hefei 230088, China}
\ead{zhangjun@ustc.edu.cn}
\vspace{12pt}

\begin{abstract}

Single-photon detectors (SPDs) are widely used in applications requiring extremely weak light detection. In the near-infrared region, SPDs based on InGaAs/InP single-photon avalanche diodes (SPADs) are the primary candidates for practical applications because of their small size, low cost and ease of operation. Driven by the escalating demands for quantum communication and lidar, the performance of InGaAs/InP SPDs has been continuously enhanced. This paper provides a comprehensive review of advances in InGaAs/InP SPDs over the past 10 years, including the investigation into SPAD structures and mechanisms, as well as emerging readout techniques for both gated and free-running mode SPDs. In addition, future prospects are also summarised.

\end{abstract}
\end{flushleft}
\vspace{12pc}
\submitto{\MST}
%
%
%
\hrulefill
\ioptwocol
\section{Introduction}
Single-photon detectors (SPDs) are key components in numerous applications, such as quantum information\cite{Rev21QP}, light detection and ranging (lidar)\cite{Rev23SPDforlidar}, and fluorescence lifetime imaging\cite{Rev20FLIM}. For near-infrared single-photon detection, the mainstream devices include superconducting nanowire single-photon detectors (SNSPDs)\cite{Rev21SNSPD,SNSPD20PDE}, up-conversion single-photon detectors (UCSPDs)\cite{ZQ19UCSPD}, and semiconductor SPDs\cite{ZJ15review}. SNSPDs exhibit desirable performance characteristics, such as high photon detection efficiency (PDE), low dark count rate (DCR), low timing jitter, and no afterpulse. However, the cryogenic operating conditions lead to a relatively large size and high costs. UCSPDs exhibit moderate performance and size, however, some intrinsic disadvantages, such as the requirement of a pump laser, narrow spectral response, and environmental sensitivity, limit their use in practical applications. In contrast, semiconductor SPDs have the advantages of small size, low cost, and ease of operation, which make them the most appropriate candidates for practical applications.

Semiconductor SPDs consist of single-photon avalanche diodes (SPADs) and readout circuits. SPADs are a special type of avalanche photodiodes (APDs) dedicated to Geiger-mode operation. In this working mode, the reverse bias voltage is higher than the breakdown voltage. Once a single photon is absorbed, the generated carries have opportunity to trigger a self-sustaining avalanche current with sufficient amplitude for identification. Readout circuits are used to discriminate the avalanche current while quenching it for subsequent photon detection. Currently, several types of SPADs can be used for near-infrared single-photon detection, including but not limited to InGaAs/InP\cite{ZJ15review}, InGaAs/InAlAs\cite{InAlAs07}, and Ge-on-Si SPADs\cite{GeSi19NC}. Among these devices, InGaAs/InP SPADs exhibit the best overall performance and are currently the most widely used devices\cite{Rev22SPAD}.

Investigation of InGaAs/InP-based single-photon detection began in the mid-1990s. Initially, commercially available InGaAs/InP APDs designed for optical communication applications were operated in the Geiger mode. However, owing to the high dark count rate, the detector must be cooled to below 100 K and operated in gated mode\cite{Lacaita96}. In the 2000s, driven by the rapid development of quantum communication, InGaAs/InP SPADs optimised for Geiger mode operations began to appear\cite{PW06,JXD07}, and the negative feedback avalanche diodes (NFADs)\cite{JXD09NFAD} dedicated to free-running operation were invented. Meanwhile, primary avalanche readout techniques, such as sine-wave gating (SWG)\cite{Nihon06SWG}, capacitance-balancing technique\cite{Tomita02}, and self-differencing\cite{YZL07SD}, were developed during this period. However, InGaAs/InP SPDs still suffered from low photon detection efficiency and high noise. Until the beginning of 2010s, the typical PDE of the InGaAs/InP SPD was $\sim$10\%\cite{ZJ12SWG}.

In the past decade, research on InGaAs/InP detectors has primarily focused on improving the performance metrics, particularly PDE and DCR. Fundamentally, significant progress has been achieved in optimising InGaAs/InP single-photon avalanche diodes (SPADs) through advancements in structural design and fabrication techniques. The physical mechanism of the InGaAs/InP SPAD was studied in detail. For synchronous photon detection applications such as quantum communication, the parameters of high-frequency gating SPD have been fully optimised, while novel gating readout circuits have been proposed to enable higher frequency and enhanced integration. For asynchronous photon detection, several methodologies have been demonstrated to achieve free-running operation, including NFADs, active quenching circuits, and gated-free operation.

In this review, we introduce the advances in InGaAs/InP SPDs over the last 10 years. The discussion begins with a brief introduction to the InGaAs/InP SPAD structure, followed by the investigations and optimisations of the InGaAs/InP SPAD characteristics. Subsequently, the recent state-of-the-art InGaAs/InP SPDs and their applications are summarised. Finally, we conclude this paper and discuss future perspectives. We hope this review will help readers quickly understand the recent developments in InGaAs/InP SPDs and capture future trends and opportunities.

Previous reviews\cite{ZJ15review,ZHP14Rev} on InGaAs/InP SPDs are recommended to provide an overview of the basic concepts and techniques. Other relevant reviews concerning quantum photonics\cite{Rev21QP}, single-photon imaging\cite{Rev23SPDforlidar}, quantum communication and quantum computation\cite{Chen21Rev}, lidar\cite{SGMJ22Rev}, and InGaAs/InP SPAD arrays\cite{Rev23Array} have also been recommended for readers as references.

\begin{figure*}[htbp]
\centerline{\includegraphics[width=16 cm]{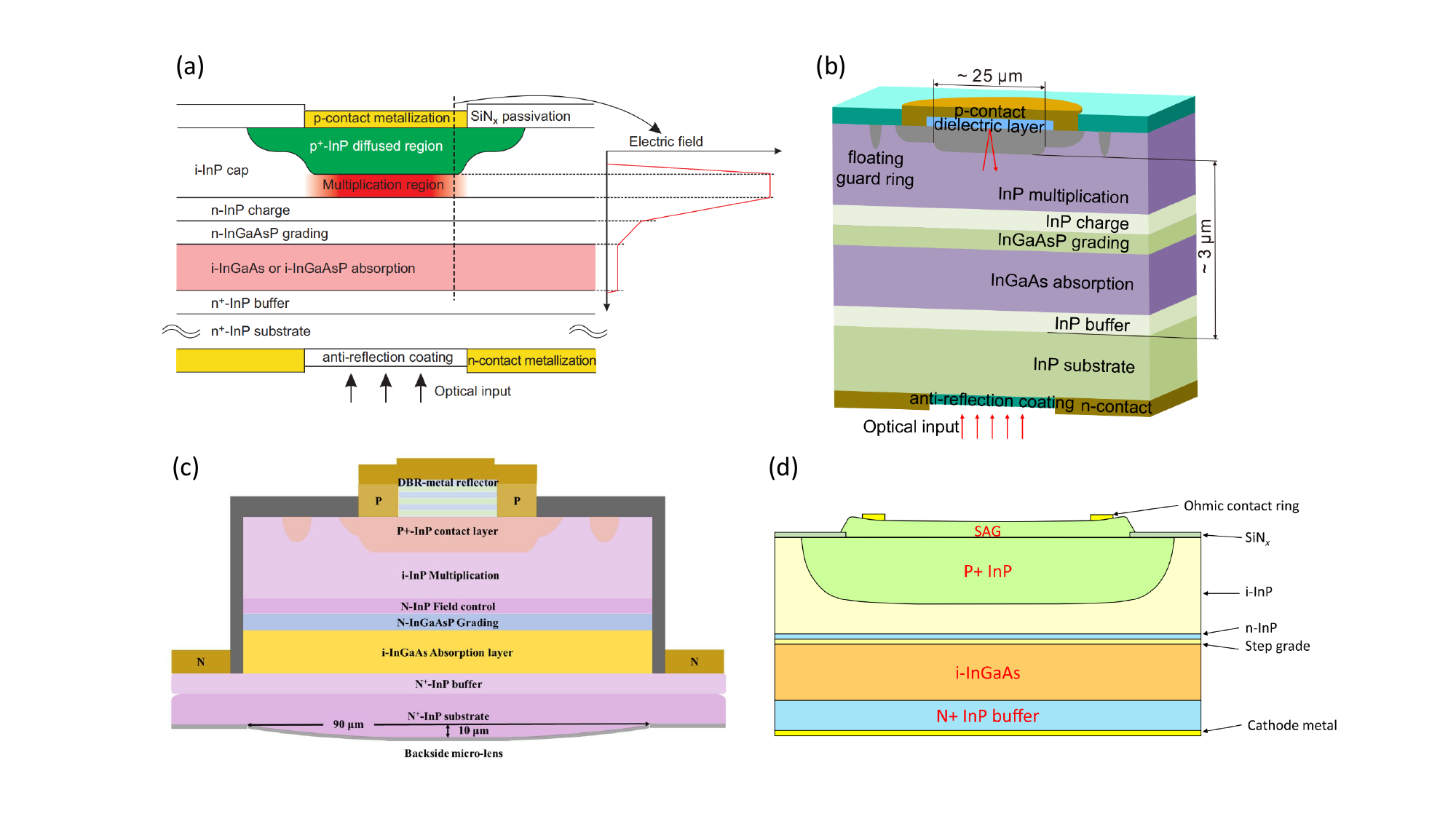}}
\caption{Semiconductor structural design of InGaAs/InP SPADs. (a) Basic structure of SAGCM. Reprinted from \cite{ZJ15review}, with permission from Springer Nature. (b) SAGCM SPAD with an additional dielectric-metal reflection layer. Reprinted from \cite{ZJ20PDE}, with the permission of AIP Publishing. (c) SAGCM SPAD with a DBR-metal reflector and a micro-lens. \textcircled{c} [2022] IEEE. Reprinted, with permission, from \cite{WL22structure}. (d) Novel structure constructed using the selective area growth (SAG) technique. Reprinted from \cite{Kiz22Stru}. CC BY 4.0.}
\label{Fig1}
\end{figure*}

\section{InGaAs/InP SPADs}

\subsection{Structure} \label{structure}

Presently, most InGaAs/InP SPADs are designed based on separate absorption, grading, charge, and multiplication (SAGCM) structures\cite{ZJ15review}, as shown in Fig.~\ref{Fig1}(a). In such a structure, an InGaAs layer with a 0.75 eV band gap is used as the absorption layer, while an InP layer with a 1.34 eV band gap is used as the multiplication layer. The electric fields in both layers are controlled by adjusting the doping concentration and thickness of the charge layer. A high electric field is designed in the multiplication layer to provide high avalanche probability, while a moderate electric field is designed in the absorption layer to guarantee a saturated drift rate of the photogenerated carriers while avoiding significant tunnelling effects. An InGaAsP grading layer is used to provide a smooth transition in the valence band, which helps to avoid carrier accumulation at the heterojunction interface. A ladder structure PN junction is usually designed using double-diffusion technology to create a uniform electric field in the central zone while avoiding a high electric field at the edge. Since incident photons must pass through the InP substrate before arriving at the absorption layer, the spectral response of the InGaAs/InP SPAD is limited to the range of 0.9 to 1.65 $\mu$m.

Recently, several optimised SPAD structures have been designed to improve the PDE while avoiding a significant increase in the DCR. As shown in Fig.~\ref{Fig1}(b), Fang $et$ $al.$\cite{ZJ20PDE} fabricated a dielectric-metal reflection layer on top of the photosensitive region, which can reflect the transmitted photons to the absorption layer with 95\% reflectivity. Therefore, the absorption efficiency was enhanced by $\sim$20\% without increasing the DCR. With this structure, the authors achieved a record PDE of 60\% at 1550 nm. For practical use, given a 3 kcps DCR, the PDE reached $\sim$40\%. Similarly, Zhang $et$ $al.$\cite{WL22structure} designed a Bragg reflector together with a microlens, as shown in Fig.~\ref{Fig1}(c), which increased the absorption efficiency by 58\%. Kizilkan $et$ $al.$\cite{Kiz22Stru} presented a novel SPAD structure using a selective area growth (SAG) technique. Based on the SAG, the thickness of the InP layer can be controlled to gradually increase from the centre to the edge, as shown in Fig.~\ref{Fig1}(d). As a result, the electric field decreases at the edge of the device, which successfully prevents edge breakdown without the need for a guard ring or a double diffusion technique. This device achieved a PDE of 43\% at 1550 nm.

\subsection{Characteristics}

The primary parameters used to characterise the performance of the InGaAs/InP SPDs include the PDE, DCR, afterpulse probability, maximum count rate, and timing jitter. These parameters influence each other, thus the optimisation of one parameter often involves performance tradeoffs with other parameters. In this section, we will introduce the mechanisms and optimisations of these parameters.

The PDE is defined as the probability that an SPD induces an output signal in response to an incident photon. For a SAGCM structure SPAD, PDE can be estimated as $PDE=\eta_{c}\times\eta_{inj}\times\eta_{abs}\times\eta_{ava}$, where $\eta_{c}$ is the coupling efficiency, $\eta_{abs}$ is the absorption efficiency, $\eta_{inj}$ is the injection efficiency from the absorption layer to the multiplication layer, and $\eta_{ava}$ is the avalanche probability. For fiber coupling devices, $\eta_{c}$ often exceeds 90\% by the design of micro-lens and anti-reflection coating~\cite{ZJ23PDE}. $\eta_{inj}$ is always considered to be approximately 100\% due to the high electric field in the active region. For a given wavelength, $\eta_{abs}$ is determined from the thickness of the absorption layer. For example, Signorelli $et$ $al.$\cite{Milan21PDE} and He $et$ $al.$\cite{He22HPDE} thickened the absorption layers to 2 $\mu$m and 2.2 $\mu$m and achieved 50\% and 55.4\% PDE at 1550 nm, respectively. $\eta_{ava}$ is primarily influenced by the electric field in the multiplication layer, thus the PDE can be enhanced by increasing the excess bias voltage.

The DCR is defined as the count rate without illumination. It is originally generated by thermal excitation, tunnelling excitation and trap-assisted tunnelling excitation\cite{Don06design}. Decreasing the defect density is a fundamental approach to improving DCR performance, however, it is dependent on the current maturity of the InGaAs material system, which makes it difficult to achieve a breakthrough improvement in a short time. From the perspective of the SPAD structure design, the DCR is strongly influenced by the volume of the active region and electric field distribution. Reducing the device diameter and thickness of the absorption layer helps improve the DCR performance. For a given SPAD device, the DCR is primarily influenced by the bias voltage and temperature. In practice, cooling SPAD devices and operating them under relatively low excess bias voltages are effective methods for realising extremely low DCR\cite{Kri14,Geneva14LDCR,Covi15FR}.

Tradeoffs exist in the optimisation of the PDE and DCR. For example, thickening the absorption layer or increasing the excess bias voltage can enhance the PDE, however, it will result in a larger DCR. Shrinking the size of the SPAD is an effective approach for reducing the DCR, but it may induce the deterioration in the coupling efficiency. Nevertheless, these two parameters can be optimised simultaneously using a novel structural design, as mentioned in Section ~\ref{structure}. Theoretical simulation methods have also been proposed\cite{Milan13design,ZJ16design,XZY19design,Croatia16design,XLZ20design}. By establishing appropriate ideal models, the PDE and DCR can be calculated using structural parameters at different excess voltages and temperatures. These simulations and detailed trends help optimise the SPAD structural design for dedicated applications. Moreover, the optimisation of fabrication technologies has also helped improve the PDE and DCR performances. For example, Fabio $et$ $al.$\cite{Milan13fabricate} optimised the growth and double p-type Zinc diffusion at different wafer temperatures and precursor gas flows. Based on this, Alberto $et$ $al.$\cite{Milan14highperf} reduces the DCR of their InGaAs/InP SPAD from 100 kcps to less than 10 kcps at a PDE of 30\%.

Afterpulse is another important parameter of InGaAs/InP SPADs, that originates from the subsequent release of trapped carriers generated by a previous avalanche. It has been experimentally proven that the afterpulse effect exerts negative impacts on various applications, such as increasing the quantum bit error rate in quantum key distribution (QKD)\cite{HZF18AP}, destroying the randomness of a quantum random number\cite{HZF15QRNG}, and distorting lidar signal\cite{ZJ17lidar}. Generally, the $P_{ap}$ can be roughly modelled as $P_{ap}(t)\propto C \int_{0}^{\Delta t}V_{ex}(t)dt \times e^{-t/\tau}$\cite{ZJ15review}, where $C$ is the equivalent capacitance, $\Delta t$ represents the avalanche duration time, and $\tau$ is the lifetime of trapped carriers. The model indicates that $Pap$ is not only determined by the SPAD characteristics, such as defects density and structural design, but is also influenced by the readout circuits. For example, high-frequency gating and fast active quenching readout circuits suppress $Pap$ by reducing the avalanche duration time $\Delta t$. The NFAD devices minimise the parasitic capacitance $C$ to suppress $Pap$.

\begin{figure}[htbp]
\centerline{\includegraphics[width=7.5 cm]{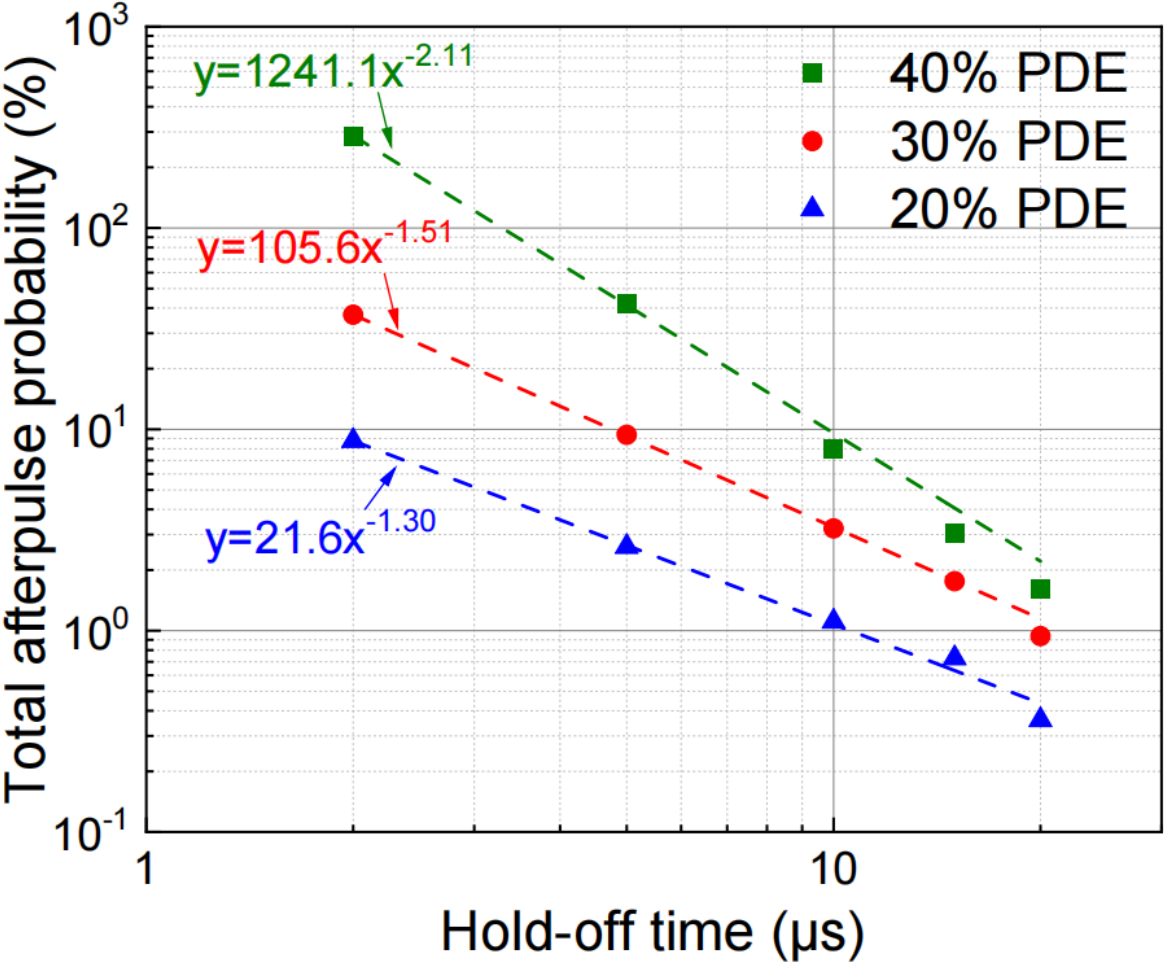}}
\caption{Power law of afterpulse probability versus hold-off time. \textcircled{c} [2024] IEEE. Reprinted, with permission, from \cite{ZJ23PDE}.}
\label{Fig2}
\end{figure}

Investigation of the intrinsic mechanism of the afterpulse effect is critical for the further optimisation of $Pap$. Recently, several hypotheses have been proposed based on experimental observations. Zhang $et$ $al.$\cite{ZJ09ASIC} reported that $P_{ap}$ does not decay exponentially with the delay time, as described above. They proposed a multiple-trapping model with two or three detrapping time to fit the experimental results. Anti $et$ $al.$\cite{Milan11afterpulse} measured the temperature dependence of the detrapping time, and evaluated the activation energies of the detects as 81.6 meV, 65.8 meV and nearly 0 meV respectively. Moreover, this model has also been used in other studies\cite{Wie16AP,HZF16AP,ZJ17lidar}. However, Itzler $et$ $al.$\cite{PLI12AP} reported that the detrapping time fitted by a multiple-trapping model has no physical significance, because the extracted values of the detrapping time depend entirely on the selected range of the data. Subsequently, they proposed a simple power law, $P_{ap}(t) = C T^{-\alpha}$ with only two fitting parameters. Worldwide research has demonstrated that the power law fits experimental data well\cite{Cova04AP,Jensen06AP,Geneva15AP,ZJ23PDE}. For example, recently published afterpulse data versus hold-off time are shown in Fig.~\ref{Fig2}\cite{ZJ23PDE}. This can be explained by the broad and continuous distribution of trap levels in the InP multiplication layer. By investigating the detrapping time over a wide temperature range, Korzh $et$ $al.$\cite{Geneva15AP} reported that the activation energies of the defects are in the 0.05 to 0.22 eV region. However, the physical significance of the parameter $\alpha$ has not yet been well explained.

In addition, Wang $et$ $al.$\cite{HZF16AP} demonstrated that the afterpulse effect is non-Markovian, implying that $P_{ap}$ at a certain moment is influenced not only by the pre-ignition avalanche, but also by all previous avalanches. This is because during an avalanche process, the number of captured carriers is considerably smaller than the total number of defects. This result can help in the development of a precise afterpulse correction algorithm for practical applications.

The maximum count rate $C_{max}$ is obtained under saturated illumination. $C_{max}$ and the DCR determine the dynamic range of the InGaAs/InP SPD. For high-frequency gating SPD, the maximum $C_{max}$ has reached as high as 500 Mcps\cite{ZHP23MCR}. However, for a free-running SPD, an additional hold-off time is usually required to suppress the afterpulse probability, which limits the enhancement of $C_{max}$. Therefore, the optimisation of $C_{max}$ essentially requires a further reduction in the afterpulse probability.

\begin{figure*}[t]
\centerline{\includegraphics[width=17 cm]{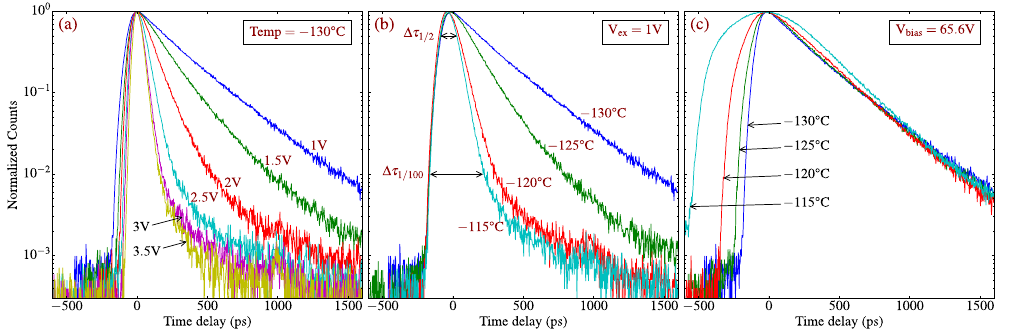}}
\caption{Temporal responses of the InGaAs/InP SPD under different conditions. Reprinted with permission from \cite{Geneva16jitter} \textcircled{c} Optical Society of America. (a) Timing jitter with different excess bias voltages at -130$^{\circ}$C. (b) Timing jitter at different temperatures with 1 V excess bias voltage. (c) The bias voltage is maintained constant and the temperature is varying.}
\label{Fig3}
\end{figure*}

Timing jitter is defined as the time uncertainty between the incident photons and output electric signals. In high-precision time-correlated single-photon counting applications\cite{ZJ2045km,ZJ21NLOS}, the temporal distribution of the SPD output signal is assumed to be Gaussian. However, the data measured by Amri $et$ $al.$\cite{Geneva16jitter} negate this assumption, as depicted in Fig.~\ref{Fig3}. In fact, the timing jitter of InGaAs/InP SPADs is attributed to two factors: transit time from the absorption layer to the multiplication layer, and the avalanche build-up time in the multiplication layer. The temporal response of the avalanche build-up process is Gaussian, while the transit process can be considered as thermionic emission, resulting in an exponential tail. Therefore, the final temporal response can be described by a convolution of Gaussian and exponential distributions.

Fig.~\ref{Fig3} (a) shows the timing jitter measured with different excess bias voltages at -130$^{\circ}$C. With the increase of electric field, both the rising and falling time decrease, corresponding to a shorter build-up time and transit time. Fig.~\ref{Fig3} (b) shows the timing jitter measured at different temperatures with 1 V excess bias voltage. The rising time is almost constant, which proves that the build-up time is only determined by the excess bias voltage. Since the breakdown voltage drops with the decrease of temperature, the absolute bias voltage drops correspondingly in the experiment, leading to a slower falling edge at low temperature. Fig.~\ref{Fig3} (c) shows the timing jitter measured at different temperatures with a constant bias voltage. Under this condition, the falling edge is almost the same, which proves the transit time is only related to the absolute bias voltage. Obviously, the rising edge become faster at low temperature due to the rise of excess bias voltage.

\begin{figure}[htbp]
\centerline{\includegraphics[width=7.5 cm]{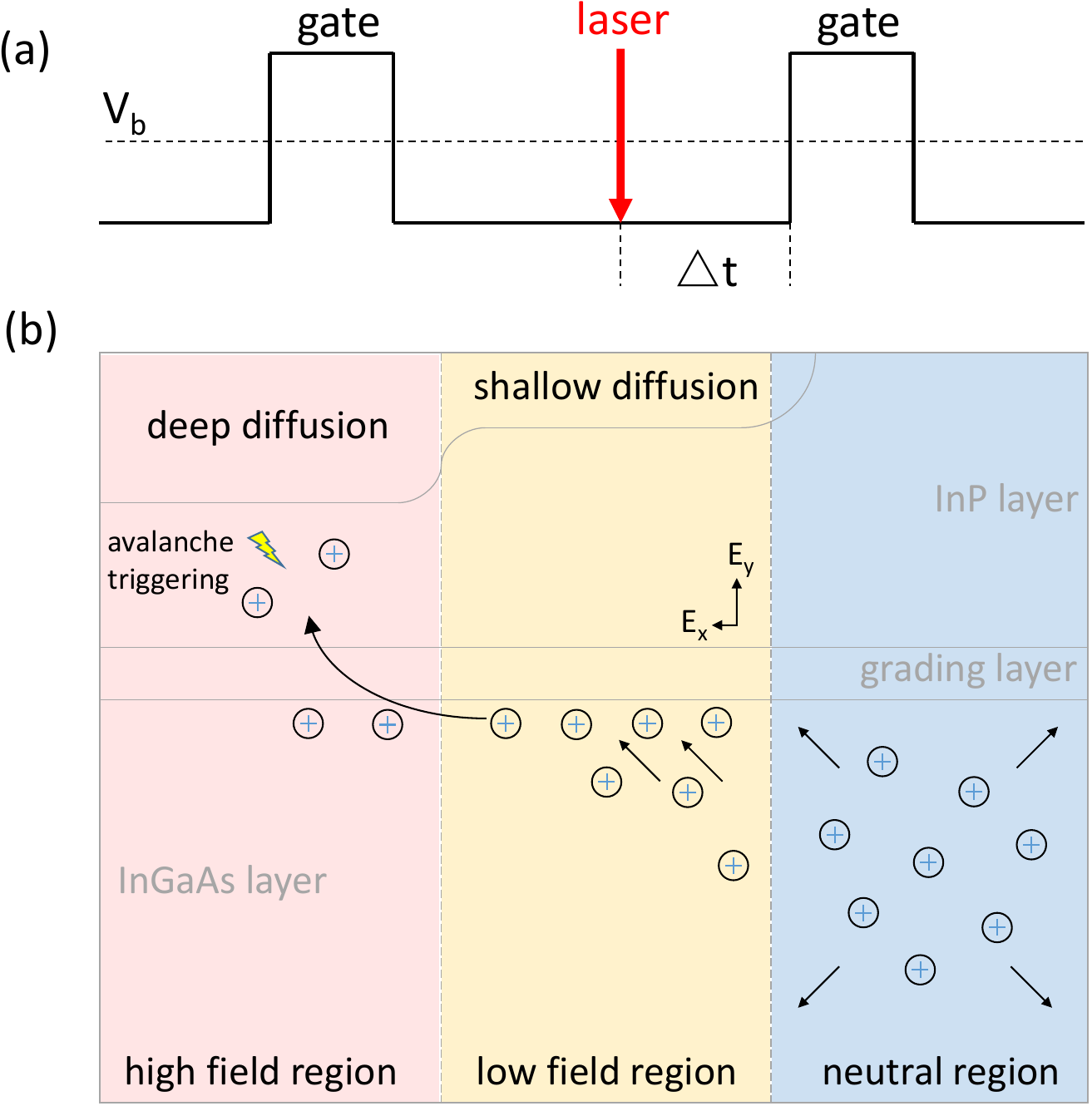}}
\caption{(a) Scheme for charge persistence measurement with a laser pulse reaching before the gate window. $V_{b}$: breakdown voltage. (b) Scheme of charge persistence mechanism.}
\label{Fig4}
\end{figure}

From the aspect of SPAD design, timing jitter can be optimised by thinning the multiplication layer or optimising the band structure of the grading layer. For a given SPAD, operating the SPAD in high-PDE mode with sufficient excess bias voltage can effectively decrease the timing jitter. Presently, the reported minimum full width at half maximum (FWHM) timing jitter of InGaAs/InP SPD is $\sim$50 ps. For example, Amri $et$ $al.$\cite{Geneva16jitter} and Xu $et$ $al.$\cite{ZJ23PDE} achieved 52 ps and 49 ps FWHM timing jitter, respectively, using a NFAD-based free-running SPD. He $et$ $al.$\cite{HZF232G5} demonstrated a high-frequency gating SPD with 44 ps FWHM timing jitter.

\begin{figure*}[hbp]
\centerline{\includegraphics[width=17 cm]{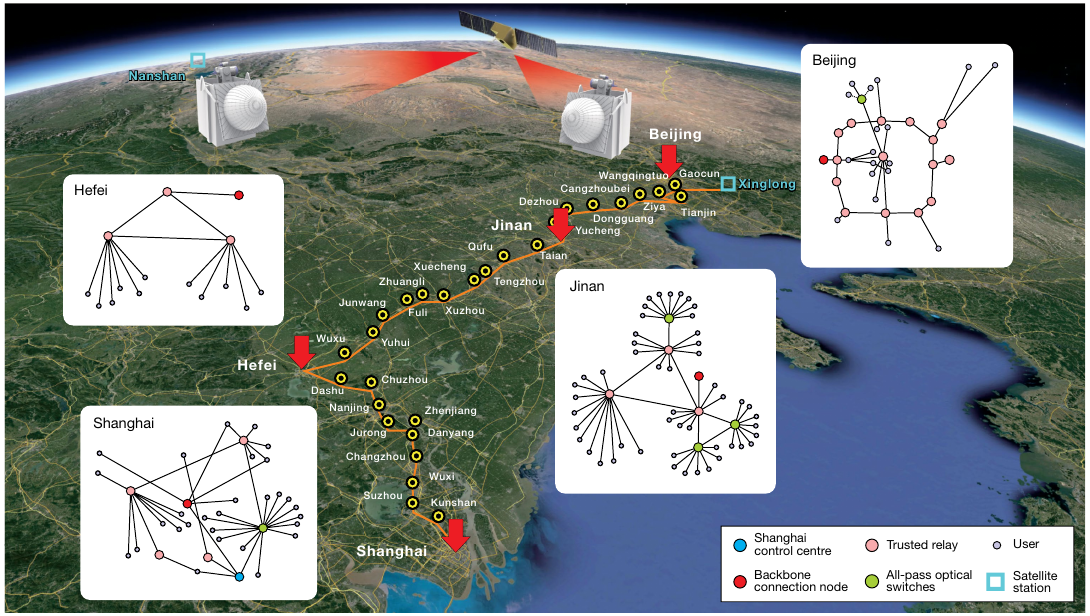}}
\caption{High-frequency sine-wave gating InGaAs/InP SPDs serve as key components of a 4600 km space-to-ground quantum communication network. The network consists of a backbone fiber link over 2000 km and two ground satellite links separated by 2600 km. Reprinted from \cite{Micius21}, with permission from Springer Nature.}
\label{Fig5}
\end{figure*}

\subsection{Charge persistence}

Charge persistence (CP) is another source of noise for InGaAs/InP SPADs\cite{ZJ09ASIC,TW18CP}. As shown in Fig.~\ref{Fig4} (a), when the photons arrive before a gate, there is a reasonable probability of triggering an avalanche during the subsequent gate-on interval. This phenomenon has long been overlooked since photons normally arrive during the gate-on time for synchronous detection. However, it plays an important role in lidar applications, in which a strong local stray laser could illuminate the SPDs during the gate-off time.

The experimental results reported by Calandri $et$ $al.$\cite{Milan16CP} demonstrated that $\eta_{off}$, the detection efficiency for photons incident during the gate-off time, exponentially increased with a decrease in temperature. Meanwhile, $\eta_{off}$ decreased with the time interval $\Delta t$ from the laser pulse to the gate edge, following a power law. In addition, they found that $\eta_{off}$ increased when the laser spot moved to the periphery of the device. Then, the authors proposed a physical model to explain their experimental results. As demonstrated in Fig.~\ref{Fig4} (b), the electric field in the InGaAs/InP SPAD is simulated according to its electrode position and doping distribution, which can be divided into high-field, low-field, and neutral regions. Photon-generated holes in the neutral region are freely diffused in all directions, and some of them are collected in the low-field region. Here, holes accumulated beneath the grading layer owing to the low tunnelling probability, and drift to high-field region driven by the horizontal component of the electric field. In the high-field region, holes tunnel into multiplication layer and eventually trigger an avalanche.

Based on this model, Telesca $et$ $al.$\cite{Milan23CP} optimised the shape and depth of the shallow zinc diffusion in InGaAs/InP SPADs to mitigate the charge persistence effect. Considering that the variation trends of the CP versus temperature and $\Delta t$ are quite similar to the afterpulse probability, this model could inspire a novel method to explain the mechanism of the afterpulse effect. Moreover, the model indicates that the thermally generated carriers outside the active region contribute to DCR, therefore, the DCR can be reduced by optimising the shallow zinc diffusion region.

\section{InGaAs/InP SPDs}

\subsection{Gated mode}

The gated mode is an effective approach for suppressing DCR due to its small duty cycle. In general, the gating techniques are divided into low-frequency (\textless 100 MHz) and high-frequency (\textgreater 100 MHz) gating. Since the low-frequency gating technique is relatively simple to implement, it has been widely used for the fundamental characterisation of InGaAs/InP SPAD\cite{Milan21PDE,He22HPDE,Bouzid14,WWJ20,Tavares22,Bin22}. A critical technique in the gating electronics is the cancellation of capacitive response signals. For low-frequency gating, the effective approaches include the coincidence method\cite{coin04}, radio frequency delay line scheme\cite{RFdelay00}, and the double-SPAD technique\cite{DSPAD02}. For the detailed implementation of these readout circuits, please refer to\cite{ZJ15review}.

Over the past decade, QKD has been the key driver for enhancing the gating frequency. There are two benefits to using high-frequency gating SPD in QKD systems. On the one hand, high-frequency gating helps improve the raw key rates. On the other hand, the avalanche current can be quickly quenched by a narrow gate, which effectively suppresses the afterpulse probability. Currently, the primary techniques for realising high-frequency gating include self-differencing\cite{YZL15}, capacitance-balancing\cite{Milan151G3gate,Park19DA,Rarity21SWG}, harmonic subtraction\cite{Res13,Res16}, and sine-wave gating (SWG)\cite{ZJ121G25}. Among these, SWG is the most appropriate candidate for practical QKD systems owing to its stability and ease of implementation. To date, high-frequency SWG InGaAs/InP SPDs have served as key components in numerous practical QKD systems\cite{Rev22QKD}, particularly, in a 4600 km space-to-ground quantum communication network\cite{Micius21} as displayed in Fig.~\ref{Fig5}.

\begin{figure}[htbp]
\centerline{\includegraphics[width=7.5 cm]{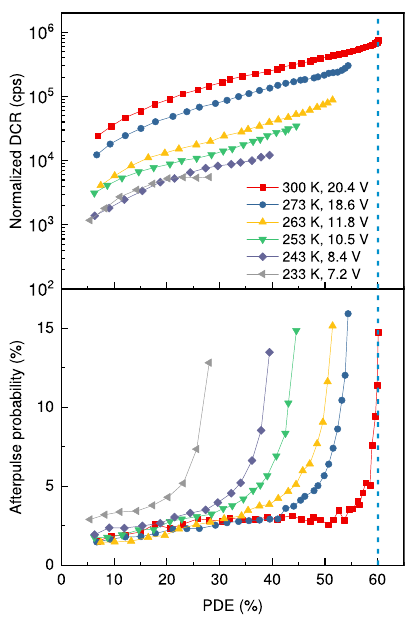}}
\caption{Normalised DCR and afterpulse probability vs PDE of a 1.25 GHz SWG SPD, as demonstrated by Fang $et$ $al.$\cite{ZJ20PDE}. The label inside represents the operating temperature and optimised gating amplitude. Reprinted from \cite{ZJ20PDE}, with the permission of AIP Publishing.}
\label{Fig6}
\end{figure}

In recent years, several approaches have been proposed to improve the performance of SWG SPDs.
For the PDE and DCR optimization, a shortened gate width has been found to benefit to improve performance\cite{ZJ20PDE,YZL15,Los22SWG}. Given a fixed gating frequency, a shortened gate width can be achieved by increasing the peak-peak amplitude (Vpp) of the SWG. However, Acerbi $et$ $al.$\cite{Milan13DCR} and Losev $et$ $al.$\cite{Los22SWG} demonstrated that the DCR increased significantly when the gate-off voltage was too low, limiting the infinite increase in the gate amplitude. For example, Tada $et$ $al.$\cite{Tada16Vpp} demonstrated an SWG SPD with an amplitude of 50 Vpp at 1.27 GHz, and achieved a PDE of 53.4\% and a DCR of 3.5$\times$10$^{-4}$/gate. In addition, it was observed that the maximum achievable PDE increases at high temperatures\cite{ZJ20PDE,YZL15}. This can be explained by the fact that the bandgap of the InGaAs absorption layers narrows at high temperatures, which leads to an increase in the absorption efficiency. By comprehensively optimising the InGaAs/InP SPAD structure, gating amplitude, and operating temperature, Fang $et$ $al.$ achieved a record PDE of 60\%, and the experimental results are depicted in Fig.~\ref{Fig6}.

Several new readout techniques have been developed for afterpulse suppression. Zavodilenko $et$ $al.$\cite{Zav17} and Losev $et$ $al.$\cite{Los22AR} demonstrated an active reset module in a 312.5 MHz SWG SPD. Fan $et$ $al.$\cite{YZL23} demonstrated an ultra-narrowband interference circuit (UNIC) and achieved a 1\% $Pap$ with a $\sim$1.4 ns dead time and a 21.2 \% PDE. He $et$ $al.$\cite{HZF17AP} reported that most afterpulses originate from successive avalanche signals and the distortion of the electrical filters. The afterpulse probability was significantly reduced by removing the wider filtered avalanche signals.

Currently, the majority of high-frequency SWG SPDs operated at a gating frequency of approximately 1 GHz. To develop the next-generation high speed QKD systems, further improvements in the gating frequency are urgently needed. SWG SPDs with gating frequencies greater than 2 GHz were demonstrated in the early 2010s\cite{ZJ102G23,Namekata102G,YZL103G}, however, further enhancement of the gating frequency is limited by the high afterpulse probability, high timing jitter, and electrical bandwidth of the SPAD devices. Liang $et$ $al.$\cite{ZHP19TRR,ZHP22TRR} investigated the performance of SPD in a wide gate frequency range from 100 MHz to 2.75 GHz, and reported that the afterpulse sharply increases when the gating frequency reaches 2.5 GHz.

\begin{figure}[htbp]
\centerline{\includegraphics[width=7.5 cm]{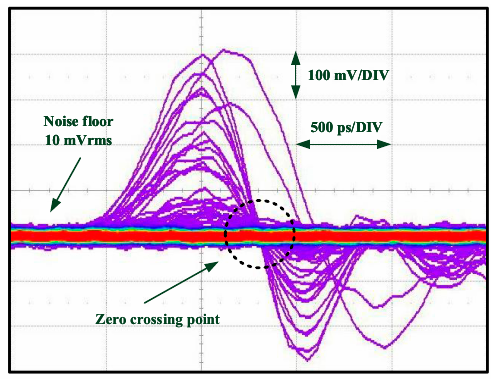}}
\caption{Filtered and amplified avalanche signals of a 2.5 GHz gated SPD. The signals have a constant zero-crossing point. Reprinted from \cite{HZF232G5}. CC BY 4.0.}
\label{Fig7}
\end{figure}

Recently, He $et$ $al.$\cite{HZF232G5} presented a practical 2.5 GHz SWG SPD. In that study, the authors claimed that filter distortion and avalanche amplitude variation were the major sources of timing jitter. The avalanche signals obtained after filtering and amplification are presented in Fig.~\ref{Fig7}. The authors noted that avalanche signals with different amplitudes had the same zero crossing point since the avalanches were quenched at the same gate-off edge. Therefore, a zero-crossing discriminator was applied to the SPD system for signal discrimination. Combined with the afterpulse suppression method mentioned previously\cite{HZF17AP}, they achieved a 44 ps timing jitter and 1.4\% $Pap$ with a 21\% PDE.

\begin{figure}[htbp]
\centerline{\includegraphics[width=7.5 cm]{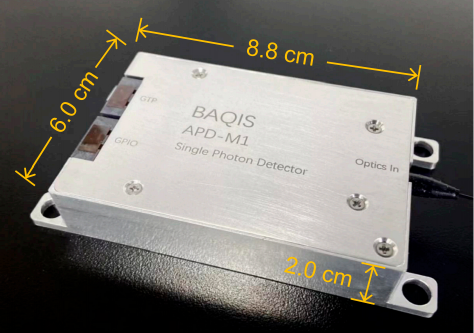}}
\caption{Compact 1.25 Ghz sine-wave-gate single-photon detector module developed by Yan $et$ $al.$\cite{YZL23Compact}. Reprinted from \cite{YZL23Compact}. CC BY 4.0.}
\label{Fig8}
\end{figure}

\begin{figure*}[t]
\centerline{\includegraphics[width=14 cm]{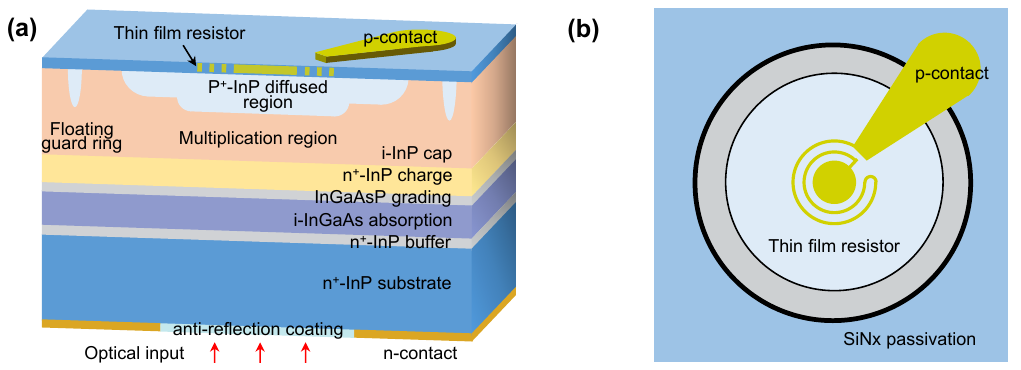}}
\caption{(a)Semiconductor structure of NFAD device with SAGCM SPAD and thin film resistor. (b) Top view of NFAD. \textcircled{c} [2024] IEEE. Reprinted, with permission, from \cite{ZJ23PDE}.}
\label{Fig9}
\end{figure*}

The integration of SPDs is also important for next-generation QKD systems. Efforts to improve the integration of SPDs have focused on three main aspects: the temperature control systems, front-gate generation circuits, and readout circuits. Liang $et$ $al.$\cite{ZHP17RT} and Liu $et$ $al.$\cite{ZHP19HPDE} demonstrated high-performance room-temperature SWG SPDs that significantly reduced the power and the size of the cooling systems. Jiang $et$ $al.$\cite{ZJ17LTCC} designed a monolithically integrated readout circuit (MIRC) with a size of 15 mm $\times$ 15 mm. Based on the MIRC chip and a butterfly package InGaAs/InP SPAD, they designed a miniaturised SWG SPD with a size of 13 cm $\times$ 8 cm $\times$ 4 cm\cite{ZJ18MSWG}. Recently, Ribezzo $et$ $al.$\cite{Milan23SIP} presented a system-in-package fast-gated InGaAs SPD, and applied it to a 100 km submarine fiber QKD system. Yan $et$ $al.$\cite{YZL23Compact} developed a compact 1.25 GHz SWG SPD module with a size of 8.8 $\times$ 6 $\times$ 2 $cm^{3}$. The photo of the module is shown in Fig.~\ref{Fig8}.

\subsection{Free-running mode}

For asynchronous photon detection, the InGaAs/InP SPDs should be operated in free-running mode. Passive quenching is a fundamental approach for realising free-running mode SPDs. However, the large quenching resistance and stray capacitance lead to a long recovery time and severe afterpulse effect. To reduce the stray capacitance, Liu $et$ $al.$\cite{WL23PQ} integrated a passive quenching resistor via wire bonding, and achieved a 3.96\% $Pap$ at a 20\% PDE with a 5 $\mu$s hold-off time. Nevertheless, the primary approach to implementing free-running InGaAs/InP SPDs over the last decade has been the use of NFADs. An NFAD device monolithically integrates a high-resistance thin-film resistor\cite{JXD09NFAD} or a zinc-diffused resistor\cite{Milan16NFAD} on the surface of an InGaAs/InP SPAD. Due to the monolithic integration, the parasitic parameters are minimised, which significantly suppresses afterpulse effect. A typical structure of an NFAD is illustrated in Fig.~\ref{Fig9}.

Decreasing the operating temperature of the NFAD is an effective method to suppress DCR\cite{Covi15FR}. Korzh $et$ $al.$\cite{Geneva14LDCR} presented a free-running InGaAs/InP SPD with only 1 cps DCR at 10\% PDE by cooling the NFAD to a temperatures of -110 $^{\circ}$C. Assisted by the low-noise SPD, they implemented a 307 km QKD\cite{Geneva15QKD} and a 2.5 GHz QKD\cite{Geneva18QKD} experiments. Yu $et$ $al.$\cite{ZJ17lidar,ZJ18lidar} demonstrated the application of a free-running InGaAs/InP SPD to accurate lidar systems. In 2021, Li $et$ $al.$\cite{ZJ21imaging} and Wu $et$ $al.$\cite{ZJ21NLOS} realised the longest 3D imaging and non-line-of-sight imaging using NFAD-based SPD. Recently, Xu $et$ $al.$\cite{ZJ23PDE} comprehensively optimised the NFAD design and operation condition, and achieved an excellent overall performances with a 40\% PDE and a 2.3 kcps DCR. The application of free-running InGaAs/InP SPD to singlet-oxygen luminescence detection has also been demonstrated\cite{Mos21O2,Geneva15O2}.

\begin{figure}[htbp]
\centerline{\includegraphics[width=7.5 cm]{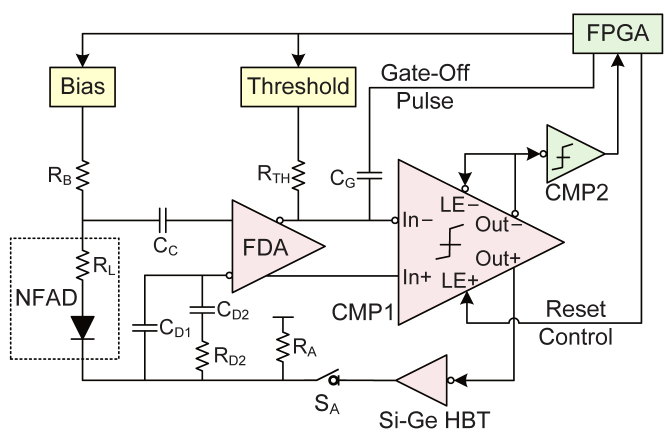}}
\caption{Schematic of active-quenching circuit. Reprinted with permission from \cite{LYF21AQ} \textcircled{c} Optical Society of America.}
\label{Fig10}
\end{figure}

The active quenching technique has also been developed for free-running operation. Acerbi $et$ $al.$\cite{Milan13AQ} demonstrated that a fast active quenching circuit reduced $Pap$ by four times compared with a simple passive quenching circuit. In 2016, Liu $et$ $al.$\cite{LYF16AQ} designed a fast active-quenching circuit with a 1.6 ns avalanche duration. In 2017, they reduced the propagational delay of the feedback loop by 375 ps, resulting in an afterpulse reduction from 10\% to less than 3\% under the conditions of a 20\% PDE and 10 $\mu$s hold-off time\cite{LYF17AQ}. To improve the maximum count rate, in 2020, they further optimised the width of the avalanche signal to less than 500 ps, and achieved a dead time as short as 35 ns with 10\% PDE and 11.6\% Pap\cite{LYF20AQ}. In 2021, they applied a fast active quenching circuit to NFAD, and reduced the total $Pap$ by $\sim$70\%\cite{LYF21AQ}. A schematic of the active quenching circuit is depicted in Fig.~\ref{Fig10}.

High-frequency gating SPDs can also be used for asynchronous photon detection, which is called gated-free detection, with an extremely high count rate at the cost of PDE loss and an increase in jitter. Tosi $et$ $al.$\cite{Milan13GateFree} demonstrated gated-free detection using a 915 MHz SWG SPD, and achieved a maximum count rate of 100 Mcps with 3\% PDE and 0.3\% $Pap$. Kong $et$ $al.$\cite{ZHP14imaging} and Li $et$ $al.$\cite{ZHP15CAM} demonstrated the application of a 1.5 GHz gated-free SPD in a 3D imaging system and a chirped amplitude modulation lidar, respectively. In 2023, Liang $et$ $al.$\cite{ZHP23MCR} presented a 1 GHz gated-free SPD with a maximum count rate of 500 Mcps. Hagihara $et$ $al.$\cite{Jap20imaging} used a gated-free SPD to demonstrate compressive single-pixel imaging.

In addition, several algorithms have been developed to correct errors caused by the imperfects in SPDs. Georgieva $et$ $al.$\cite{Georgieva21} demonstrated an analytical model to describe the relationship between incident photon numbers and the output count rate considering dark counts and dead time. Yu $et$ $al.$\cite{ZJ17lidar} developed an afterpulsing correction algorithm for lidar applications. The corrected signals agreed very well with the results obtained using the SNSPD, with a relative error of $\sim$2\%. For cases where the incident photon intensity varies rapidly, Yu $et$ $al.$\cite{ZJ18lidar} proposed a dynamic hold-off time correction algorithm. These algorithms facilitate the use of InGaAs/InP SPDs for high accuracy applications.

\begin{table*}[htp]
\centering
\caption{Performances of InGaAs/InP SPADs/SPDs reported over the last decade.}
\label{t1}
\renewcommand\arraystretch{2}
\begin{tabular}{|m{0.8cm}<{\centering}|m{0.8cm}<{\centering}|m{4cm}<{\centering}|m{1.2cm}<{\centering}|m{1.6cm}<{\centering}|m{1.2cm}<{\centering}|m{1cm}<{\centering}|m{1cm}<{\centering}|m{1cm}<{\centering}|}
\hline
\hline
Ref                       & Year             & Readout Technique                      & Temp               & PDE$^a$          & DCR (cps)$^b$      &$T_{h}$              & $Pap$             & Jitter (ps)        \\
\hline
~\cite{Bouzid14}          & 2014             & 10 MHz gate, 1 ns width                &-40$^{\circ}$C      & 20.1\%           & 120                &/                    & 1.2\%             &/                   \\
~\cite{Geneva14LDCR}      & 2014             & NFAD                                   &-110$^{\circ}$C     & 27.7\%           & 15.2               &20 $\mu$s            & $\sim$20\%        &129                 \\
~\cite{Milan151G3gate}    & 2015             & 1.3 GHz SWG                            &-30$^{\circ}$C      & 30\%             & 28.6 k             &1.5 ns               & 1.5\%             &70                   \\
~\cite{YZL15}             & 2015             & 1 GHz self-difference                  &20$^{\circ}$C       & 55\%             & $\sim$300 k        &10 ns                & 10.2\%            &91                   \\
~\cite{ZHP17RT}           & 2017             & 1.5 GHz SWG                            &20$^{\circ}$C       & 21\%             & 120 k              &80 ns                & 1.4\%             &82                   \\
~\cite{Park19DA}          & 2019             & 1 GHz capacitance-balancing            &-20$^{\circ}$C      & 20.4\%           & 21.7 k             &160 ns               & 3.5\%             &/                    \\
~\cite{ZHP19HPDE}         & 2019             & 1 GHz SWG                              &21$^{\circ}$C       & 50.4\% (1310 nm) & 310 k              &10 ns                & 5.6\%             &/                   \\
~\cite{Tada20HPDE}        & 2020             & 1.27 GHz SWG                           &16$^{\circ}$C       & 55.9\%           & 597 k              &200 ns               & 4.8\%             &150                  \\
~\cite{WWJ20}             & 2020             & 1 MHz gate, 4 ns width                 &-47$^{\circ}$C      & 70\% (1064 nm)   & 48 k               & /                   &/                  &/                   \\
~\cite{ZJ20PDE}           & 2020             & 1.25 GHz SWG                           &27$^{\circ}$C       & 60\%             & 340 k              &88 ns                & 14.8\%            &/                   \\
~\cite{LYF21AQ}           & 2021             & Active quench                          &-50$^{\circ}$C      & 10\%             & 918                &80ns                 & 20.4\%            &/                   \\
~\cite{Milan21PDE}        & 2021             & 1 MHz gate, 2 ns width                 &-48$^{\circ}$C      & 50\%             & 20 k               &5 $\mu$s             & 4\%               &70                   \\
~\cite{Bin22}             & 2022             & 1 MHz gate, 4.4 ns width               &-40$^{\circ}$C      & 20\%             & 320                &1 $\mu$s             & 0.57\%            &/                   \\
~\cite{He22HPDE}          & 2022             & 1 MHz gate, 12.5 ns width              &-26$^{\circ}$C      & 55.4\%           & 43.8               &/                    & /                 &/                   \\
~\cite{WL22structure}     & 2022             & 50 MHz gate, 1 ns width                &-40$^{\circ}$C      & 30\%             & 665                &200 ns               & 15\%              &/                   \\
~\cite{HZF232G5}          & 2023             & 2.5 GHz SWG                            &/                   & 21\%             & 3.75 k             &20 ns                & 1.4\%             &44                   \\
~\cite{YZL23}             & 2023             & 1.25 GHz SWG                           &-30$^{\circ}$C      & 30\%             & 1.6 k              &2 ns                 & 2.3\%             &/                   \\
~\cite{ZJ23PDE}           & 2023             & NFAD                                   &-60$^{\circ}$C      & 40\%             & 2.3 k              &10 $\mu$s            & 8 \%              &49                   \\

\hline
\hline
\end{tabular}
\begin{flushleft}
\ \ \ \  Temp: operation temperature, PDE: photon detection efficiency, DCR: dark count rate, $T_{h}$: hold-off time \\
\ \ \ \  $Pap$: afterpulse probability, SWG: sing-wave gating, NFAD: negative-feedback avalanche diode. \\
\ \ \ \  $^a$ The PDE values without declaration are measured at 1550 nm.\\
\ \ \ \  $^b$ DCR here represents the actual dark counts per second rather than the normalized DCR.
\end{flushleft}
\end{table*}

\section{Conclusions and outlook}
The InGaAs/InP SPDs are the most practical candidates for near-infrared single-photon detection. In the past decade, the SPAD structural design, fabrication technology, and material quality have greatly improved, resulting in a significant performance enhancements. In particular, the PDE improved from a typical value of 10\% to a maximum of 60\% at 1550 nm. The readout and affiliated circuits of both the gating and free-running modes have been continuously optimised to achieve the limited performance of SPAD devices. Several state-of-the-art InGaAs/InP SPADs or SPDs presented in the last decade are listed in Table.~\ref{t1} for readers as references. Currently, high-performance InGaAs/InP SPDs are commercially available\cite{QCD600B,ID_Qube,PDMIR,SPD_A_NIR}, and serve as key components in numerous applications, such as the QKD systems, single-photon imaging, gas sensing, and aerosol lidars.

Although significant improvements have been achieved in the last decade, the noise in InGaAs/InP SPDs is still relatively large compared to that in Si SPDs. In the future, optimisations of the noise characteristics of InGaAs/InP SPDs will be the most urgent task. For this purpose, optimizing material quality and fabrication technology are crucial steps that require sustained long-term efforts. Additionally, further enhancements can be made to the semiconductor structure. For instance, by precisely controlling the electric field distribution of the InGaAs/InP SPAD, the probability of free carriers drifting into the active region can be reduced, thereby mitigating DCR. Advanced fiber coupling technology combined with smaller size InGaAs/InP SPADs also facilitate the development of low-noise SPDs. For afterpulsing suppression, the current methods, such as high-speed gating, NFAD, and active quenching, focus on reducing the total number of avalanche carriers. Future strategies might explore optical or physical methods to regulate the lifetime of trapped carriers, aiming to further decrease the afterpulse probability. Furthermore, the physical mechanisms of the afterpulse effect and charge persistence should be studied in more detail to guide the design of low-noise SPAD devices. To meet the requirements of next-generation high-speed QKD systems, InGaAs/InP SPDs should be operated in the ultrahigh-frequency gating mode, which requires optimisations of both SPAD devices and readout circuits for higher bandwidth and lower timing jitter. Finally, the development of integrated readout circuits and system-in-package SPDs are also critical tasks for practical applications.

\section*{Acknowledgments}
This work is supported by the Innovation Program for Quantum Science and Technology (2021ZD0300804) and the National Natural Science Foundation of China (62175227).

\section*{Data availability statement}
No new data were created or analysed in this study.

\section*{References}

\bibliography{SPD}
\bibliographystyle{iopart-num}

\end{document}